\documentclass[letterpaper,11pt]{article}
\usepackage{jheppub}
\usepackage{amsmath}
\usepackage{physics}
\usepackage{relsize}
\usepackage{tikz}
\usepackage{pgfplots}
\usepackage{comment}

\usepackage[utf8]{inputenc}

\title{The role of dRGT mass terms in cutoff holography and the Randall--Sundrum II scenario}
\author[a]{Nicholas Ondo\footnote{Now at Google LLC; all work done prior to current employment.}}
\author[b]{Vasudev Shyam}
\affiliation[b]{Stanford Institute for Theoretical Physics and Department of Physics, Stanford University, Stanford, CA 94305, USA}
\emailAdd{nicholas.ao.ondo@gmail.com}
\emailAdd{vshyam@stanford.edu}

\begin{document}
\abstract{\\
    We show how the $T^2$ deformation of large $N$, holographic conformal field theories arise from coupling them to the quadratic mass term of ghost-free massive gravity. We then show that in a certain approximation, the equations arising from the variation of the background frame field are those of the Randall--Sundrum II scenario.}
\maketitle
\section{Introduction}
\label{sec:Intro}
One of the major challenges in Quantum Gravity is understanding holographic observables in a finite volume.  The discovery of AdS/CFT allows for a complete understanding of Quantum Gravity in principle, but in its original form, it only readily provides a description of gravitational observables in spaces with an asymptotically AdS boundary.  How the correspondence works when the asymptotic boundary is replaced by a Dirichlet wall bounding a finite volume is far from well understood. A full accounting is expected to aid in exporting holography to non-AdS spacetimes, even those that feature in our description of Cosmology \cite{dSdSTTbar,Flauger:2022hie}.  Recent progress has been made on holography in finite-volume with the advent of the $T\bar{T}$ deformation \cite{Smirnov:2016lqw, Cavaglia:2016oda} of the boundary CFT theory, both for $AdS_3$ quantum gravity theories \cite{McGough:2016lol,klm2018,Mazenc:2019cfg} and its generalization to higher dimensions dubbed the $T^2$ deformation, for large N CFTs \cite{Taylor:2018xcy, Hartman:2018tkw, Caputa:2019pam}. 

The basic picture is that the field theory dual to gravity in the region of finite spacetime volume bounded by the Dirichlet wall is the holographic CFT deformed by these irrelevant operators. Furthermore, the flow equation that these deformed theories satisfy map to the equation satisfied by the gravitational action on shell when the radial location of the Dirichlet wall is varied. In this sense, the $T\bar{T}$ or $T^2$ deformations \emph{move the CFT into the bulk} \cite{McGough:2016lol, Hartman:2018tkw,Taylor:2018xcy}. 

In most studies of these solvable irrelevant operator deformations, the observables of these theories are defined implicitly through a flow equation.
However, it is profitable to have an explicit representation of the partition function of the deformed theory itself. In the case of the two dimensional $T\bar{T}$ deformation, it was shown in \cite{Freidel:2008sh,Tolley:2019nmm, McGough:2016lol, Mazenc:2019cfg} that there is a kernel with which the partition function of the seed theory can be convolved with to produce the partition function of the deformed theory. This kernel was recognized by Tolley in \cite{Tolley:2019nmm} to be the action of two dimensional ghost free massive gravity. In this article, we show how the higher dimensional $T^2$ deformation can also be generated by coupling the generating function of large N holographic CFTs to the quadratic mass term of ghost free massive gravity. This is our main result. 

The work of Randall and Sundrum in \cite{Randall_1999} provides another perspective on gravity in AdS space. They show how gravity can be localized on a brane of co-dimension one in AdS space, provided this surface has a certain tension and is the fixed point locus of a $\mathbb{Z}_2$ action. In other words, they show that if we took two patches of AdS space where one is the $\mathbb{Z}_2$ mirror copy of the other and join them along a co-dimension 1 brane with a certain tension, then junction conditions dictate that the Einstein equations for the induced metric on this surface are satisfied with zero cosmological constant, and in the presence of a matter stress tensor localized only on the brane \cite{Shiromizu_2000}. This remarkable mechanism is the prototypical example of how gravity can be localized on a surface of lower dimension while the extra dimension remains infinitely large! 

The interplay between the $AdS/CFT$ correspondence and the RS II scenario described above was first studied in \cite{Arkani_Hamed_2001}, where they dub the $\mathbb{Z}_2$ fixed point brane the ``Planck brane". They argue that on this brane resides a cutoff version of the dual CFT, now also coupled to lower dimensional gravity. In other words, the holographic CFT with a cutoff provides the matter that couples to gravity localized on the brane. We will show that the cutoff of the CFT is provided by the $T^2$ deformation and that the brane world Einstein equations can be derived from our holographic formalism when applied to the configuration resulting from gluing two $\mathbb{Z}_{2}$ mirrored patches of radially cutoff $AdS$. 

\subsection*{Organization of this article}
\begin{itemize}
    \item In section \ref{sec:fle} we derive a general expression for the deformation of the action on shell in AdS space truncated at a finite radius by a Dirichlet wall. 

    \item In section \ref{sec:cutoff} we use the heuristics of dimensional deconstruction to guide us to a form of effective action of the the holographic dual theory living on the cutoff surface. We also show that this effective action satisfies the flow equation derived in the previous section.

    \item Then in \ref{sec:branewrld} we show how a linearization of the deformed holographic CFT action provides the right matter stress tensor for gravity localized on the Planck brane in the context of the Randall Sundrum II scenario. 

    \item Finally, section \ref{sec:conclusions} contains our conclusions and some future outlook. 
\end{itemize}

\subsection*{Added note}
An article \cite{TateoConti_2022} exploring a similar geometry based approach to defining the $T^2$ deformation of classical field theories in $d>2$ appeared while the present article was in preparation. The approach here is focused on the $T^2$ deformation of large $N$ holographic CFTs as opposed to classical field theories that \cite{TateoConti_2022} focuses on. 

\section{The flow equation for the cutoff action on shell}
\label{sec:fle}
In this section, we will derive an equation that the gravitational action for a region in $AdS$ space bound by a finite Dirichlet wall satisfies when the location of the wall is varied. 

To start, we need a useful representation of the action on shell. We start with the Einstein Hilbert action with Gibbons--Hawking term and counterterms: 
\begin{equation}
   S  =  \int_{M} d^{D+1} x \sqrt{-g} \left(R - \frac{D(D-1)}{\ell^2}\right)  - 2 \int_{\partial M}d^D x \sqrt{-h}\textrm{tr}K + S_{c.t.}
\end{equation}
Here $M$ denotes spacetime, $\partial M$ it's boundary, $\ell$ denotes the $AdS$ radius, $h_{\mu\nu}$ is the metric induced on the boundary, $K_{\mu\nu}=\nabla_{(\mu}n_{\nu)}$ is the extrinsic curvature and $n_{\mu}$ is the normal to $\partial M$.

Following steps similar to \cite{Hawking_1996, Banihashemi:2022jys}, we write 
\begin{equation}
    R = -2(G_{\mu\nu} - R_{\mu\nu})n^{\mu}n^{\nu}.
\end{equation}

Then within the action:
\begin{equation*}
    S = \int_M d^{D+1} x \sqrt{-g}\bigg(-2(G_{\mu\nu}-R_{\mu\nu})n^{\mu}n^{\nu} +\end{equation*} 
    \begin{equation}- \frac{D(D-1)}{\ell^2} g_{\mu\nu}n^{\mu}n^{\nu}\bigg)+ 2 \int_{\partial M} d^{D}x \sqrt{-h}
    \textrm{tr}K  + S_{c.t.}.
\end{equation}
\begin{equation*}
    => S = -2\int_M d^{D+1}x \sqrt{-g} \bigg[\bigg( G_{\mu\nu} + \frac{(D(D-1))}{2\ell^2}g_{\mu\nu}\bigg)n^{\mu}n^{\nu}- \end{equation*}
    \begin{equation}
    -R_{\mu\nu}n^{\mu}n^{\nu} \bigg] + 2 \int_{\partial M} d^{D}x \sqrt{-h}\textrm{tr}K +S_{c.t.}.
\end{equation}

Then, to go on shell we use the equations of motion 
\begin{equation}
    G_{\mu\nu} = -\frac{D(D-1)}{2\ell^2}g_{\mu\nu},
\end{equation}

which when contracted twice with the normal vectors $n^{\mu}$ give us the radial ADM Hamiltonian constraint \cite{Arnowitt:1962hi}. 

We denote the action on shell as $\mathcal{S}$: 
\begin{equation}
    \mathcal{S} = 2\int_{M} d^{D+1}x \sqrt{-g} R_{\mu\nu}n^{\mu}n^{\nu} + 2 \int \sqrt{-h}\textrm{tr}K+S_{c.t.}.
\end{equation}

Then we use the identity: 
\begin{equation*}
    R_{\mu\nu}n^{\mu}n^{\nu} = n^{\mu}[\nabla_{\mu},\nabla_{\nu}]n^{\nu} = \bigg((\nabla_{\mu}n^{\mu})(\nabla_{\nu}n^{\nu}) - (\nabla_{\mu}n^{\nu})(\nabla_{\nu}n^{\mu}) -\end{equation*} \begin{equation}- \nabla_{\mu}(n^{\mu}\nabla_{\nu}n^{\nu})+ \nabla_{\nu}(n^{\mu}\nabla_{\mu}n^{\nu})\bigg) 
\end{equation}
\begin{equation}
    = \textrm{tr}K^2 - K^{\mu}_{\nu}K^{\nu}_{\mu}-\nabla_{\mu}(n^{\mu}\textrm{tr}K) + \nabla_{\nu}(n^{\mu}\nabla_{\mu}n^{\nu}).
\end{equation}
Now, plugging this back into the action and using Stokes' theorem\footnote{First, $\int d^{D+1}x \sqrt{-g}(\nabla_{\mu}(n^{\mu}\textrm{tr}K)) = \int d^Dx \sqrt{-h}\textrm{tr}K$, which cancels against the GHY term, and $\int d^{D+1}x \sqrt{-g}(\nabla_{\nu}(n^{\mu}\nabla_{\mu} n^{\nu})) = \int d^Dx \sqrt{-h}n_{\nu}\nabla_{\mu}n^{\nu} = 0$ because $n_{\nu}n^{\nu}=1$.} we find 
\begin{equation}
    \mathcal{S} =-2\int d^{D+1}x \sqrt{-g}(K^{\mu}_{\nu}K^{\nu}_{\mu}-\textrm{tr}K^2) +S_{c.t.}.
\end{equation}
We then choose the following decomposition of the bulk metric with radial lapse $N$ and shift $N_{\mu}$:
\begin{equation}
    ds^2 = N^2d r^2 + N_{\mu}dx^\mu dr + r^2 h_{\mu\nu}dx^{\mu}dx^{\nu},
\end{equation}
which then implies $\sqrt{-g} = Nr^D\sqrt{-h}$. Note that the Dirichlet wall is placed at $r= r_c$. We finally obtain 
\begin{equation}
    \mathcal{S} = -2 \int ^{r_c}_{0}dr \int d^{D}x  N r^D\sqrt{-h} (K^{\mu}_{\nu}K^{\nu}_{\mu}-\textrm{tr}K^2)+ S_{c.t.}.
\end{equation}

Then we find 
\begin{equation}
     \partial_{r_c}\mathcal{S}-\partial_{r_c}S_{c.t.}=  -2r^D_c \int_{\partial M} d^{D}x N \sqrt{-h}(K^{\mu}_{\nu}K^{\nu}_{\mu}-\textrm{tr}K^2).
\end{equation}
This equation is typically written in terms of the Brown York \cite{BY,Balasubramanian_1999} stress tensor: 
\begin{equation}
    T^{\mu\nu}_{BY}=(K^{\mu\nu} - \textrm{tr}K h^{\mu\nu})-\frac{1}{\sqrt{-h}}\frac{\delta S_{c.t.}}{\delta h_{\mu\nu}}.
\end{equation}

in terms of which we have 
\begin{equation*}
    \partial_{r_c}\mathcal{S} - \partial_{r_c}S_{c.t.} = -2r^D_c \int_{\partial M}d^DxN \sqrt{-h}\left(h_{\mu(\rho}h_{\sigma)\nu} - \frac{1}{D-1}h_{\mu\nu}h_{\rho\sigma}\right)\times
\end{equation*}
\begin{equation}
    \bigg(T^{\mu\nu}_{BY}-\frac{1}{\sqrt{-h}}\frac{\delta S_{c.t.}}{\delta h_{\mu\nu}}\bigg) \bigg(T^{\rho\sigma}_{BY}-\frac{1}{\sqrt{-h}}\frac{\delta S_{c.t.}}{\delta h_{\rho\sigma}}\bigg).
\end{equation}

When interpreted through the lens of the holographic duality, the Brown York stress tensor is identified with the stress tensor of the dual deformed quantum field theory living on the boundary. Then, the scale radius duality identifies the radial direction in the bulk with the renormalization group scale of the dual field theory. This in turn means that the above equation, when translated through the holographic dictionary is the RG flow equation for the dual field theory describing the deformation of the holographic CFT by the so called $T^2$ operator.

For ease of notation, moving forward we will denote 
\begin{equation}
    K_{\mu\nu\rho\sigma} = \left(h_{\mu(\rho}h_{\sigma)\nu} - \frac{1}{D-1}h_{\mu\nu}h_{\rho\sigma}\right).
\end{equation}

To  put this flow equation into the form found in \cite{Hartman:2018tkw}, where the authors considered $D=3,4$, we take $N=1$ and note that the counterterms are given by a 0 derivative and 2 derivative piece: 
\begin{equation}
    S_{c.t.} = S^{(0)}_{c.t.}+S^{(2)}_{c.t.},
\end{equation}
where 
\begin{equation}
    S^{(0)}_{c.t.}  = \frac{(D-1)}{\ell}\int d^D x \sqrt{-h} 
\end{equation}
\begin{equation}
    S^{(2)}_{c.t.} = -\frac{\ell}{2(D-2)}\int d^Dx \sqrt{-h} R(h).
\end{equation}
Then, from noting that on $\partial M$, $r_c$ plays the role of the conformal factor meaning that for any functional of the boundary metric $F[h\vert_{\partial M} = r^2_ch]$:
\begin{equation}
    \partial_{r_c}F = 2\int_{\partial M}d^D x \sqrt{-h}h_{\mu\nu}\frac{\delta F}{\delta h_{\mu\nu}},
\end{equation}
and the conditions:
\begin{equation}
    \partial_{r_c}S^{(0)}_{c.t.}= 2\int d^{D}x\frac{1}{\sqrt{-h}} K_{\mu\nu\rho\sigma}\frac{\delta S^{(0)}_{c.t.}}{\delta h_{\mu\nu}} \frac{\delta S^{(0)}_{c.t.}}{\delta h_{\rho\sigma}},
\end{equation}
\begin{equation}
    \partial_{r_{c}}S^{(2)}_{c.t.} = 4\int d^{D}x \frac{1}{\sqrt{-h}} K_{\mu\nu\rho\sigma}\frac{\delta S^{(0)}_{c.t.}}{\delta h_{\mu\nu}}\frac{\delta S^{(2)}_{c.t.}}{\delta h_{\rho\sigma}},
\end{equation}
we have the flow equation: 
\begin{equation}
    \partial_{r_c} \mathcal{S}= \int d^{D}x \sqrt{-h}K_{\mu\nu\rho\sigma} \left(T^{\mu\nu}_{BY}-\frac{1}{\sqrt{-h}}\frac{\delta S^{(2)}_{c.t}}{\delta h_{\mu\nu}}\right)\left(T^{\rho\sigma}_{BY}-\frac{1}{\sqrt{-h}}\frac{\delta S^{(2)}_{c.t}}{\delta h_{\rho\sigma}}\right).
\end{equation}
which agrees with the $D=3,4$ results in \cite{Hartman:2018tkw,Belin:2020oib,Caputa:2019pam}.

In the next section, we will show coupling the dRGT mass term to the large N holographic CFT dual to gravity in the case with the asymptotic boundary leads to an effective action satisfying the above flow equation interpreted in field theory terms.

\section{Defining the theory living on the cutoff surface}
\label{sec:cutoff}
The key entry of the holographic dictionary we will use is the following: 
\begin{equation}
    \log Z_{CFT}[h] = \mathcal{S}_{ren}[h].
\end{equation}
The left hand side denotes the renormalized generating function, given by the logarithm of the partition function of the large N holographic CFT. The right hand side the finite part of the holographic action on shell in asymptotically AdS space. By finite, we are referring to the fact that without having subtracted certain counter terms, the action on shell in asymptotically AdS space diverges. 

The argument of the left hand side is the background metric describing the geometry on which the CFT lives. This metric has no dynamics of it's own - it is a background field. Stress tensor correlation functions are computed through successive functional differentiation of this object with respect to this background metric. The argument of the right hand side is the boundary value of the dynamical bulk metric. Going on shell reduces the dependence of the gravitational action to the dependence on the boundary conditions of the metric and other dynamical fields. Here we take Dirichlet boundary conditions meaning the boundary value of the metric is held fixed and as such, it plays a similar role to that of the argument of the field theory's background metric. Variations of the (holographically) renormalized action on shell with respect to this metric computes the Brown York stress tensor, through which we can compute the globally conserved energy and momentum charges of the bulk theory. 

The extrapolation of this dictionary entry to the case where the asymptotic boundary has been replaced by a finite Dirichlet wall is given by:
\begin{equation}
    \log Z[h] = \mathcal{S}[r^2_c h],
\end{equation}

where the object on the right hand side satisfies the flow equation from the previous section.

In what follows, we will give an explicit representation for $\log Z[h]$ in terms of $\log Z_{CFT}$.

\subsection{Interlude: frame fields and related structures}

Moving forward, it will help to introduce frame fields $h^a_{\mu}$ (also known as vielbein or tetrad fields) so that 
\begin{equation}
    h_{\mu\nu} = \eta_{ab}h^a_{\mu}h^{b}_{\nu}.
\end{equation}
In terms of this object, $K_{\mu\nu\rho\sigma}$ is replaced by 
\begin{equation}
    K^{ab}_{\mu\nu} = \left(h^a_{\nu}h^{b}_{\mu}-\frac{1}{D-1}h^a_{\mu}h^b_{\nu}\right).
\end{equation}
The sense in which it is like $K_{\mu\nu\rho\sigma}$ is that the following contraction of two arbitrary 2 tensors:
\begin{equation}
    K_{\mu\nu\rho\sigma}F^{\mu\nu}F^{\rho\sigma} = F^{\mu\nu}F_{\mu\nu} - \frac{1}{D-1}(F^{\sigma}_{\sigma})^2
\end{equation}
where indices on $F^{\mu\nu}$ are raised and lowered exclusively WRT $h$.  For example, the mixed internal spacetime index object is given by 
\begin{equation}
    F^{\mu}_a = F^{\mu\nu}h^{a}_{\nu}
\end{equation}
is computed through 
\begin{equation}
    K^{ab}_{\mu\nu} F^{\mu}_a F^{\nu}_b = F^{\mu}_c F^{c}_{\mu} - \frac{1}{D-1}(h^d_{\sigma}F^{\sigma}_d)^2.
\end{equation}
Also, it will help to know that the inverse of this structure is given by $f^{[\mu}_af^{\nu]}_b$, meaning:
\begin{equation}
    h^{[\mu}_{a}h^{\nu]}_b K^{bd}_{\nu\sigma} = - \delta^{\mu}_{\sigma}\delta^d_a.
\end{equation}

Another very useful quantity we will use is the generalized Kronecker delta symbols.

They are denoted
\begin{equation}
    \delta_m(X_1\cdots X_m) = \delta^{\mu_1\cdots \mu_{m}}_{\nu_1\cdots \nu_m}X^{\nu_1}_{1\,\mu_1}\cdots X^{\nu_m}_{m\,\mu_m} 
\end{equation}
where
\begin{equation}
    \delta^{\mu_1\cdots \mu_{m}}_{\nu_1\cdots \nu_m} = \delta^{[\mu_1}_{\nu_1}\cdots \delta^{\mu_m]}_{\nu_m}.
\end{equation}

Importantly, note that 
\begin{equation}
    \frac{1}{D!}\delta_{D}(h^{D}) = \det h
\end{equation}
and so 
\begin{equation}
    \frac{1}{(D-2)!}\delta_{D}\left(X_{1} X_{2} h^{D-2}\right)=\det h \delta_{2}\left(h^{-1} X_{1} h^{-1} X_{2}\right) = h^{[\mu}_a h^{\nu]}_b X^a_{1\,\mu} X^{b}_{2\,\nu}\det h.
\end{equation}

Equipped with these objects, we can very easily define the dRGT mass term to which we couple $\log Z_{CFT}$ to generate the $T^2$ deformed generating function. 

\subsection{Deconstructing the annular action on shell}
We will start by deriving our main result through an algorithm that has been used to derive the action of massive gravity theories in $D$ dimensions starting from the action for general relativity in $D+1$ dimensions. This algorithm has, in past literature been dubbed \emph{dimensional deconstruction} \cite{ArkaniHamed:2001nc,Ondo:2017kea,deRham:2013awa}. 

The aim of this exercise is to obtain a kernel that when convolved with the CFT generating function produces a solution to the flow equation derived in the previous section. In other words, to find an object $K(e,f)$ such that 
\begin{equation}
    \mathcal{S} = \min_e\left(\log Z_{CFT}[e]+K(e,f)\right).
\end{equation}


Unlike past studies of this method where it was applied directly to the Einstein-Hilbert action, we will apply it to the annular action action on shell. This object is a slight variation on the one we studied in the previous section. The difference is that in the bulk term, instead of considering the geometry between $r=0$ and the cutoff surface at $r=r_c$, we look at the annulus running between $r=r_c$ and $r= R\rightarrow \infty$, i.e. the asymptotic boundary.

We will denote this annular action $\mathcal{S}_{ann}$: 
\begin{equation}
    \mathcal{S}_{ann} = -2 \int^{R}_{r_c} dr \int d^{D}x \det h  \left(h^{[\mu}_a h^{\nu]}_b K^{a}_{\mu} K^{b}_{\nu}\right) -S_{c.t.}(r_c) + S_b(R). 
\end{equation}
Here we again tookthe lapse $N(r)=1$ and
\begin{equation}
    f^{[\mu}_a h^{\nu]}_b K^{a}_{\mu}K^{b}_{\nu} = K^{\mu}_a K^{a}_{\mu} - (h^a_{\mu}K^{\mu}_a)^2
\end{equation}
which when rewritten in terms of $K^{\nu}_{\mu} = K^{a}_{\mu}f^{\nu}_a$ gives us back the form of the on shell action in metric variables from the previous section. 

There will be boundary action terms both at the surface at $r_c$ and at the asymptotic boundary, and we will identify exactly what they are in terms of quantities in the holographically dual field theory in what follows.

In the gauge we are in, i.e. with zero shift and unit radial lapse, the extrinsic curvature can be written as 
\begin{equation}
    K^a_{\mu} = \partial_{r} h^a_{\mu},
\end{equation}

so the bulk term in the annular action can be written as 
\begin{equation}
    -\frac{2(D-1)}{(D-1)!}\int^{R}_{r_c} dr \int d^{D}x \delta_{D}((\partial_{r}h)^2h^{D-2}).
\end{equation}

The idea of dimensional deconstruction is to replace the radial derivative here by a finite difference between the frame fields evaluated at the $r=r_c$ surface, which we denote $f^a_{\mu}$ and the one on the $r=R$ surface that we denote $e^a_{\mu}$: 

\begin{equation}
     -\frac{2(D-1)}{(D-1)!}\int^{R}_{r_c} dr \int d^{D}x \delta_{D}((\partial_{r}h)^2h^{D-2})\rightarrow \frac{1}{\lambda(D-1)!}\int d^{D}x \delta_{D}((e-f)^2f^{D-2}).
\end{equation}

This is nothing but the quadratic mass term of dRGT \cite{deRham:2010ik, deRham:2010kj, Hassan:2011hr, Hassan:2011tf} massive gravity where $\frac{1}{\lambda}$ plays the role of the mass parameter, and the various numerical conversion factors are chosen for later convenience. 

Finally, adding in the boundary contributions we have in total:

\begin{equation}
    \mathcal{S}_{ann} = -\frac{1}{\lambda(D-1)!}\int d^{D}x \delta_{D}((e-f)^2f^{D-2})-S_{c.t.}[f]+ S_{b}[e].
\end{equation}

The quantity $S_{c.t.}[f]$ is given by the standard holographic counter-terms, whereas the boundary contribution at the asymptotic boundary is given by 
\begin{equation}
    S_b[e] = -\gamma\log\left(\frac{\epsilon}{\lambda}\right)\int d^{D}x \mathcal{A}(e)
\end{equation}
where $\gamma$ is a constant whose value will be fixed in the next subsection and $\mathcal{A}(e)$ is the densitized version of the conformal anomaly of the dual CFT. It is defined through the relation 
\begin{equation}
    e^a_{\mu}\frac{\delta}{\delta e^a_{\mu}}\log Z_{CFT}[e] = \mathcal{A}(e),
\end{equation}
as such, it also contains the anomaly coefficient in its definition. The parameter $\epsilon$ denotes a fixed reference scale that is needed to compensate for the fact that $\lambda$ is not dimensionless. We can interpret this scale as the cutoff of the CFT itself, and as such, generically $\lambda \gg \epsilon\ll 1.$

The annular action is defined as the difference between the action on shell in a space with an asymptotic boundary and the on shell action in the case where the boundary is at the cutoff surface $r=r_c$. Through the AdS/CFT dictionary, the former is identified with the CFT generating function. We therefore have:
\begin{eqnarray}
    \log Z_{CFT} - \mathcal{S} = \mathcal{S}_{ann}\\
    => \log Z_{CFT} - \mathcal{S}_{ann} = \mathcal{S}.
\end{eqnarray}

So we have the following representation for the cutoff action on shell:
\begin{equation*}
    \mathcal{S} = \min_e\bigg(\frac{1}{\lambda(D-1)!} \int d^{D}x \delta_{D}((e-f)^2 f^{D-2}) + \end{equation*}
    \begin{equation}+\gamma\log\left(\frac{\epsilon}{\lambda}\right)\int d^{D}x \mathcal{A}(e) + S_{c.t.}[f] + \log Z_{CFT}[e]\bigg).
\end{equation}

This expression requires slight modification in order to be correct. We will see in the following section that in order to satisfy the right flow equation, we need to add a $-\frac{1}{\lambda}\det f$ term to the above action:
\begin{equation*}
    \mathcal{S} = \min_e\bigg(\frac{1}{\lambda(D-1)!} \int d^{D}x \delta_{D}((e-f)^2 f^{D-2}) -\frac{1}{\lambda}\int d^{D}x\det f+ \end{equation*}
    \begin{equation}+\gamma\log\left(\frac{\epsilon}{\lambda}\right)\int d^{D}x \mathcal{A}(e) + S_{c.t.}[f] + \log Z_{CFT}[e]\bigg).
\end{equation}
The action $\mathcal{S}$ depends only on the geometry of the cutoff surface, i.e. the vielbein $f^a_{\mu}$, and thus we have to integrate out the field $e^a_{\mu}.$

We have therefore found 
\begin{equation*}
    K(e,f)  = \frac{1}{\lambda(D-1)!} \int d^D x \delta_{D}((e-f)^2 f^{D-2}) - \frac{1}{\lambda}\int d^D x\det f + \end{equation*}
    \begin{equation}\gamma \log \left(\frac{\epsilon}{\lambda}\right)\int d^D x \mathcal{A}(e) + S_{c.t.}[f] 
\end{equation}

The deconstruction method is more of a ansatz than a precise mechanism to generate the effective action of the dual theory, and thus we tolerate some minimal modifications like the one above. 

It is worth noting that the relationship between the deforming operator in the context of cutoff holography and the annular action in $D=2$ was noted in \cite{Caputa_2021}. 

\subsection{Reproducing the $T^2$ flow equation}
We frist write:
\begin{equation}
    \mathcal{S}[f] = \min_e \mathcal{S}[e,f],
\end{equation}
\begin{equation*}
    \mathcal{S}[e,f] = \frac{1}{\lambda(D-1)!} \int d^{D}x \delta_{D}((e-f)^2 f^{D-2}) -\frac{1}{\lambda}\int d^{D}x\det f+
\end{equation*}
\begin{equation}
+\gamma\log\left(\frac{\epsilon}{\lambda}\right)\int d^{D}x \mathcal{A}(e) + S_{c.t.}[f] + \log Z_{CFT}[e].
\end{equation}
We have an explicit representation for $\mathcal{S}[e,f]$ and an implicit one for $\mathcal{S}[f]$, but in order to show that the latter satisfies the right flow equation, we use an intermediate between the two where we subtract from $\mathcal{S}[e,f]$ a certain redundant operator. 

To see this, note that we are imposing 
\begin{equation}
    \frac{\delta \mathcal{S}[e,f]}{\delta e^a_{\mu}} = 0,
\end{equation}
which in turn also means that 
\begin{equation}
    e^a_{\mu} \frac{\delta \mathcal{S}[e,f]}{\delta e^a_{\mu}} = 0.
\end{equation}
Using the form of $\mathcal{S}[e,f]$ above, we see that this means 
\begin{equation}
    \frac{2}{\lambda(D-1)!}\delta_{D}(e(e-f)f^{D-2}) +\mathcal{A}(e)=0.
\end{equation}

Here we have used that 
\begin{equation}
    e^a_{\mu}\frac{\delta \mathcal{A}(e)}{\delta e^a_{\mu}}=0.
\end{equation}
Then, we can subtract this redundant operator from $\mathcal{S}[e,f]$ to write 
\begin{equation*}
    \mathcal{S}^* = -\frac{1}{\lambda(D-1)!}\int d^Dx \delta_{D}((e-f)f^{D-1}) - \frac{1}{\lambda}\int d^D x \det f +
\end{equation*}
\begin{equation}
    +\left(\gamma\log \left(\frac{\epsilon}{\lambda}\right)-\frac{1}{2}\right)\int d^D x \mathcal{A}(e)+ S_{c.t.}[f] + \log Z_{CFT}[e].
\end{equation}

This object is effectively a representation of $\mathcal{S}[f]$ if we replace the dependence of this object on $e^a_{\mu}$ with the on shell value $e^a_{*\mu}=e^a_{*\mu}(f)$. We write it implicitly as: 
\begin{equation}
    \mathcal{S}[f] = \mathcal{S}^*[e_*,f]=
\end{equation}
\begin{equation*}
    =-\frac{1}{\lambda(D-1)!}\int d^Dx \delta_{D}((e_*-f)f^{D-1}) - \frac{1}{\lambda}\int d^D x \det f +
\end{equation*}
\begin{equation}
    +\left(\gamma\log \left(\frac{\epsilon}{\lambda}\right)-\frac{1}{2}\right)\int d^D x \mathcal{A}(e_*)+ S_{c.t.}[f] + \log Z_{CFT}[e_*].
\end{equation}
 Moving forward, we drop the $*$ subscript for $e^a_{\mu}$ assuming implicitly that variations w.r.t. it are set to vanish. This in turn implies that when we take $f^a_{\mu}$ variations of the above object, the terms coming from the variation of $e^a_{*\mu}(f)$ with respect to $f^a_{\mu}$ will vanish. 
 
Like we mentioned in the previous section, we want $\lambda$ to play the role of $r_c$ and the latter plays the role of the conformal factor of the geometry on $\partial M$. This means we rescale 
\begin{equation}
    \tilde{f}^a_{\mu} = \lambda^{-\frac{1}{D}}f^a_{\mu},\,\, \tilde{e}^a_{\mu} = \lambda^{-\frac{1}{D}}e^a_{\mu}.
\end{equation}

Then we can write
\begin{equation*}
    \mathcal{S}[\tilde{f}] = -\frac{1}{(D-1)!}\int d^D x \delta_{D}((\tilde{e}-\tilde{f})\tilde{f}^{D-1}) - \int d^D x \det \tilde{f}+ S_{c.t.}[\tilde{f}]+
\end{equation*}
\begin{equation}
    +\left(\left(\gamma - \frac{1}{D}\right)\log\left( \frac{\epsilon}{\lambda}\right)-\frac{1}{2}\right)\int d^D x \mathcal{A}(\tilde{e}) + \log Z_{CFT}[\tilde{e}].
\end{equation}

The extra $\frac{1}{D}$ coefficient of the logarithm above comes from re-scaling the argument of $\log Z_{CFT}[e]$.

Then we compute 
\begin{equation*}
    \lambda \partial_{\lambda}\mathcal{S}[\tilde{f}] - \lambda \partial_{\lambda}S_{c.t.}[\tilde{f}] = -\frac{1}{D}\int d^D x f^a_{\mu}\frac{\delta}{\delta f^a_{\mu}}(\mathcal{S}-S_{c.t.})-\gamma\int d^Dx \mathcal{A}(e)=
\end{equation*}
\begin{equation}
   = -\frac{(D-1)}{D(D-1)!} \int d^D x \delta_{D}((\tilde{e}-\tilde{f})\tilde{f}^{D-1})-\gamma \int d^D x \mathcal{A}(e).
\end{equation}

Now we define the Brown York stress tensor here as: 
\begin{equation}
   \det \tilde{f} \tilde{T}^{\mu}_{BY\, a} = \frac{\delta \mathcal{S}}{\delta \tilde{f}^a_{\mu}} = \frac{(D-1)}{(D-1)!} \delta_{D}((\tilde{e}-\tilde{f})\tilde{f}^{D-2})^{\mu}_a + \frac{\delta S_{c.t}}{\delta \tilde{f}^a_{\mu}}. 
\end{equation}
Here we see why we had to correct the form of the action we first obtained from deconstruction in the previous subsection. We can rewrite the above expression as
\begin{equation*}
    \tilde{T}^{\mu}_{BY\,a} = \frac{1}{\det \tilde{f}}\frac{(D-1)}{(D-1)!}\delta _{D}((\tilde{e}-\tilde{f})\tilde{f}^{D-2})^{\mu}_a + \frac{1}{\det \tilde{f}} \frac{\delta S_{c.t.}}{\delta \tilde{f}^a_{\mu}}.
\end{equation*}
\begin{equation}
    =\tilde{f}^{[\mu}_a \tilde{f}^{\rho]}_d (\tilde{e}-\tilde{f})^d_{\rho} + \frac{1}{\det \tilde{f}} \frac{\delta S_{c.t.}}{\delta \tilde{f}^a_{\mu}}.
\end{equation}

Therefore 
\begin{equation}
    \tilde{T}^{\mu}_{BY\,a} -\frac{1}{\det \tilde{f}} \frac{\delta S_{c.t.}}{\delta \tilde{f}^a_{\mu}}= \tilde{f}^{[\mu}_a \tilde{f}^{\rho]}_d (\tilde{e}-\tilde{f})^d_{\rho}.
\end{equation}
Writing this in terms of un-rescaled variables: 
\begin{equation}
    T^{\mu}_{BY\,a} -\frac{1}{\det f} \frac{\delta S_{c.t.}}{\delta f^a_{\mu}}= \frac{1}{\lambda}f^{[\mu}_a f^{\rho]}_d (e-f)^d_{\rho}.
\end{equation}
the right hand side is the discretized version of the derivative $\partial_rh^a_{\mu}$, which when written in terms of the extrinsic curvature gives us the relation 
\begin{equation}
     T^{\mu}_{BY\,a} -\frac{1}{\det f} \frac{\delta S_{c.t.}}{\delta f^a_{\mu}}= f^{[\mu}_a f^{\nu]}_b K^b_{\nu},
\end{equation}
which is exactly the relationship we have in the bulk! In order for these relations to hold, we crucially needed the extra $-\frac{1}{\lambda}\det f$ term in the action. So this term is in fact needed for the self consistency of the idea of deconstruction. 

We now compute 
\begin{equation*}
    \int d^{D}x \det \tilde{f} \tilde{K}^{ab}_{\mu\nu}\left(\tilde{T}^{\mu}_{BY\,a} - \frac{1}{\det f}\frac{\delta S_{c.t.}}{\delta f^a_{\mu}}\right)\left(\tilde{T}^{\nu}_{BY\,b} - \frac{1}{\det f}\frac{\delta S_{c.t.}}{\delta f^b_{\nu}}\right) =
\end{equation*}
\begin{equation}
    = -\frac{(D-1)}{(D-1)!}\int d^D x \delta_{D}\left((\tilde{e}-\tilde{f})^2 \tilde{f}^{D-2}\right),
\end{equation}
which we simplify using the redundancy condition: 
\begin{equation}
    \frac{(D-1)}{(D-1)!}\int d^Dx \delta_{D}\left((\tilde{e}-\tilde{f}) \tilde{f}^{D-1}\right)+\frac{(D-1)}{2}\int d^D x \mathcal{A}(e).
\end{equation}

We see that if we fix
\begin{equation}
    \gamma = \frac{D-1}{2D},
\end{equation}
the following flow equation is satisfied: 
\begin{equation}
   \lambda \partial_{\lambda}\mathcal{S} - \lambda \partial_{\lambda}S_{c.t.} = -\frac{1}{D}\int d^D x \det \tilde{f}\tilde{K}^{ab}_{\mu\nu} \left(\tilde{T}^{\mu}_{BY\,a}-\frac{1}{\det \tilde{f} }\frac{\delta S_{c.t.}}{\delta \tilde{f}^a_{\mu}}\right)\left(\tilde{T}^{\nu}_{BY\,b}-\frac{1}{\det \tilde{f} }\frac{\delta S_{c.t.}}{\delta \tilde{f}^b_{\nu}}\right).
\end{equation}

We can write this in terms of the quantities without the $\tilde{()}$ as: 
\begin{equation}
     \partial_{\lambda}\mathcal{S} -  \partial_{\lambda}S_{c.t.} = -\frac{1}{D} \int d^D x K^{ab}_{\mu\nu}\left(T^{\mu}_{BY\,a} - \frac{1}{\det f} \frac{\delta S_{c.t.}}{\delta f^a_{\mu}}\right)\left(T^{\nu}_{BY\,b} - \frac{1}{\det f} \frac{\delta S_{c.t.}}{\delta f^b_{\nu}}\right).
\end{equation}

If we change variables 
\begin{equation}
    \lambda = r^{-2D}_c,
\end{equation}
we obtain
\begin{equation}
    \partial_{r_c}\mathcal{S} -  \partial_{r_c}S_{c.t.} = -2\int d^D x K^{ab}_{\mu\nu}\left(T^{\mu}_{BY\,a} - \frac{1}{\det f} \frac{\delta S_{c.t.}}{\delta f^a_{\mu}}\right)\left(T^{\nu}_{BY\,b} - \frac{1}{\det f} \frac{\delta S_{c.t.}}{\delta f^b_{\nu}}\right).
\end{equation}
This is exactly the form of the flow equation we found in the previous section. 
To recover the CFT partition function, we consider 
\begin{equation}
    \mathcal{S}-S_{c.t.}
\end{equation}
and take $\lambda = \epsilon$ along with
\begin{equation}
    e^a_{\mu} =\bar{e}^a_{\mu}= \left(1-\frac{1}{\sqrt{D}}\right)f^a_{\mu},
\end{equation}
under which
\begin{equation}
    \frac{1}{\lambda(D-1)!} \delta_{D}((\bar{e}-f)^2f^{D-2}) - \frac{1}{\lambda}\det f = 0.
\end{equation}

If we consider fluctuations around this point, 
\begin{equation}
    e^a_{\mu} = \left(1-\frac{1}{\sqrt{D}}\right)f^a_{\mu}+ \epsilon \delta h^a_{\mu},
\end{equation}
then we will have 
\begin{equation}
    \mathcal{S} - S_{c.t.} = \int d^D x \frac{\epsilon}{(D-1)!}\delta_{D}((\delta h)^2f^{D-2}) + \log Z_{CFT}[\bar{e}+ \epsilon \delta h^a_{\mu}],
\end{equation}
Recalling that $\epsilon\ll 1$, we have from the above: 
\begin{equation}
    \mathcal{S}-S_{c.t.} = \frac{\epsilon}{(D-1)!}\int d^D x \delta_{D}(\delta h^2f^{D-2}) +\log Z_{CFT}[\bar{e}(f)] + \int d^D x \det f \epsilon \delta h^a_{\mu} T^\mu_{a\,CFT}
\end{equation}
Then, minimization over $e^a_{\mu}$ in the general case corresponds to integrating out $\delta h^a_{\mu}$. Doing so leads to the following formula: 
\begin{equation}
    \mathcal{S}-S_{c.t.} = \log Z_{CFT}[\bar{e}(f)] + \frac{\epsilon(D-1)}{4} \int d^D x \det f K^{ab}_{\mu\nu}T^{\mu}_{CFT\,a}T^{\nu}_{CFT\,b}.
\end{equation}
the stress tensor computed by taking the $f^a_{\mu}$ variation of the above expression is given by: 

\begin{equation*}
    T^{\mu}_a = T^{\mu}_{CFT\,a} + \frac{\epsilon(D-1)}{2} \bigg(\bigg(T^{\mu}_{CFT\,d}f^{d}_{\sigma}T^{\sigma}_{CFT\,a}-\frac{1}{D-1}(T^{\rho}_{CFT\,d} f^{d}_{\rho})T^{\mu}_{CFT\,a}\bigg) -\end{equation*}\begin{equation} - \frac{f^\mu_a}{2}K^{cd}_{\alpha\beta}T^{\alpha}_{CFT\,c}T^{\beta}_{CFT\,d}\bigg).
\end{equation}
This expression will be important to us in the following section.

\section{Braneworld gravity}
\label{sec:branewrld}
Consider the following setup: take two patches of empty $AdS_{D+1}$ space that are mirror images of each other, then glue them along a brane of dimension $D$. Provided the brane has appropriate tension, it was shown by Randall and Sundrum in \cite{Randall_1999} that $D$ dimensional Einstein equations with zero cosmological constant are satisfied along the brane. Stress--energy localized on the brane solely sources the gravitational field. 

A natural question to ask is how to interpret this setup from the perspective of the AdS/CFT correspondence. 
An answer to this question was proposed in \cite{Arkani_Hamed_2001}, where the brane was dubbed the ``Planck brane" and the stress energy sourcing the gravitational field on the brane was identified with that of the dual CFT, cutoff in the appropriate way. The idea was that the result of having the Planck brane not be the fixed asymptotic boundary, but rather a surface at some finite distance from either mirrored patch of the bulk AdS space meant that the dual theory has to be cutoff, and more importantly, must perceive a dynamical gravitational field.
On the other hand, if we impose Dirichlet boundary conditions at some finite radius, we expect the dual theory living there to be the holographic CFT deformed by the $T^2$ operator. This begs the question: how is the Planck brane picture of RSII related to the $T^2$ deformation? We attempt to answer this question in the present section. 

In particular, we see that the precise notion of the cutoff that is appropriate to describe this physics is exactly what is provided by the $T^2$ deformation we have been studying in this article. 
\subsection{Field equations on the brane}
Following \cite{Shiromizu_2000}, we seek the gravitational field equations induced on a brane of some tension $\kappa$ that sits between two $\mathbb{Z}_2$ mirrored patches of $AdS_{D+1}.$ 

First we note that the bulk field equations, are given by
\begin{equation}
    ^{(D+1)}G^{\mu}_a = \Lambda q^{\mu}_a,
\end{equation}
where $\,^{(D+1)}G^{\mu}_a$ denotes the Einstein tensor and $q^{\mu}_a$ is the inverse of the bulk frame field. We call the frame field on the brane $f^a_{\mu}$. From the results of \cite{Shiromizu_2000}, we have that the Einstein tensor induced on the brane is given by: 
\begin{equation}
    ^{(D)}G^{\mu}_a = -\frac{(D-2)\Lambda}{D} f^{\mu}_a -E^{\mu}_a - \bigg(\left((K^{\rho}_d f^d_{\rho})K^{\mu}_a - f^{\mu}_d K^{d}_{\nu}K^{\nu}_{a} \right)-\frac{f^{\mu}_a}{2}f^{[\alpha}_df^{\beta]}_d(K^c_{\alpha}K^d_{\beta})\bigg).
\end{equation}
Here, the bulk Einstein equations, the Gauss equation and the relationship between the Riemann and Weyl tensors have been used. Here, 

\begin{equation}
    E^{\mu}_{a} = \,^{(D+1)}C^{\sigma}_{\alpha \rho \beta} n^{\alpha} f^{\mu}_{k} f^{k}_{\sigma} n^{\beta}  f^{\rho}_{a} 
\end{equation}

is the electric part of the bulk Weyl tensor. $K^{a}_\mu$ denotes the extrinsic curvature. 

The object $E^{\mu}_a$ encodes the effect of gravitational waves in the bulk. Assuming that the bulk is empty and hence contains no sources for gravitational waves lets us drop this term. Then, in order to simplify what remains, we make use of the Israel junction conditions. 

The junction conditions demand the continuity of the metric across the brane and that the jump in the extrinsic curvature: 
\begin{equation}
    [K]^{a}_\mu = K^{a}_{+\,\mu} - K^{a}_{-\,\mu}
\end{equation}

is cancelled by tension and other stress energy contributions localized on the brane. The fact that the two patches we are trying to glue are $\mathbb{Z}_{2}$ mirrored allows us to take 
\begin{equation}
    K^{a}_{-\,\mu}  = - K_{+\,\mu}.
\end{equation}

Then, if we denote the total brane stress energy as 
\begin{equation}
    S^{\mu}_{a} = -\sigma f^{\mu}_a + \tau^{\mu}_a,
\end{equation}
where 
\begin{equation}
    \sigma = (D-1)\kappa
\end{equation}
the Junction condition tells us 
\begin{equation}
    K^{a}_{+\,\mu} = - K^{a}_{-\,\mu}\equiv K^{a}_{\mu} = K^{ab}_{\mu\nu} S^{\nu}_b.
\end{equation}

Then, in the expression for the induced Einstein tensor on the brane, we have 
\begin{equation*}
    \,^{(D)}G^{\mu}_a = -\frac{(D-2)}{D}\Lambda f^{\mu}_a -\kappa^2\frac{(D-2)}{2(D-1)} f^{\mu}_a +\frac{(D-2)}{(D-1)}\kappa \tau^{\mu}_a -\end{equation*} \begin{equation}-\bigg(\left(\tau^{\mu}_{c}f^{c}_{\rho}\tau^{\rho}_{a} - \frac{1}{D-1}(f^c_{\rho}\tau^{\rho}_c)\tau^{\mu}_a\right) - \frac{1}{2}f^{\mu}_a K^{cd}_{\alpha\beta}\tau^{\alpha}_c\tau^{\beta}_d\bigg)
\end{equation}

Recalling that 
\begin{equation}
    \Lambda = \frac{D(D-1)}{2\ell^2},
\end{equation}
we find that if 
\begin{equation}
    \kappa = \frac{(D-1)}{\ell},
\end{equation}
the first two terms cancel leaving 
\begin{equation}
    ^{(D)}G^{\mu}_a = \frac{(D-2)}{\ell}\tau^{\mu}_a - \bigg(\left(\tau^{\mu}_{c}f^{c}_{\rho}\tau^{\rho}_{a} - \frac{1}{D-1}(f^c_{\rho}\tau^{\rho}_c)\tau^{\mu}_a\right) - \frac{1}{2}f^{\mu}_a K^{cd}_{\alpha\beta}\tau^{\alpha}_c\tau^{\beta}_d\bigg)
\label{eq:brane_efe}.\end{equation}

To recover Einstein's equations with zero cosmological constant and with the source being only the brane localized stress energy, the story usually goes as follows:

We take 
\begin{equation}
    \tau^{\mu}_a = c_b T^{\mu}_{brane\,a}
\end{equation}
where $T^{\mu}_{brane\,a}$ is the brane energy momentum tensor and $c_b$ is a parameter taken to be small. In particular, if we drop terms of order $c^2_b$, we are left with 
\begin{equation}
    \,^{(D)}G^{\mu}_a = \frac{(D-2)c_b}{\ell}T^{\mu}_{brane\,a}.
\end{equation}

Which is the advertised result, where $\frac{(D-2)c_b}{\ell}$ plays the role of the gravitational constant on the braneworld.

On the other hand, if we make no approximations, then the resulting stress energy tensor comes with a contribution that is quadratic in $\tau^{\mu}_a$: 
\begin{equation}
 T^{\mu}_{\textrm{total}\,a} =    \frac{(D-2)}{\ell} \tau^{\mu}_a - \bigg(\left(\tau^{\mu}_c f^{c}_{\rho}\tau^{\rho}_a - \frac{1}{D-1} (f^c_{\rho}\tau^{\rho}_c)\tau^{\mu}_a \right) - \frac{1}{2}f^{\mu}_a K^{cd}_{\alpha\beta}\tau^{\alpha}_c\tau^{\beta}_d\bigg) 
\end{equation}

This can be mapped exactly to the one arising from the small $\lambda$ limit of $\mathcal{S}-S_{ren}$ we obtained in the previous section provided we take 
\begin{equation}
    \tau^{\mu}_a = \frac{\ell}{(D-2)}T^{\mu}_{CFT\,a}
\end{equation}
and 
\begin{equation}
\epsilon = \frac{2(D-2)}{(D-1)\ell}. 
\end{equation}

In other words, if we treated $(\mathcal{S}-S_{ren})\vert_{\lambda\sim \mu \ll1}$ as the matter action on the brane, we obtain the induced field equations on the brane when we turn off the bulk Weyl tensor contribution.

In studies of braneworld cosmology (see for instance \cite{Maartens_2010} and references therein) this contribution to the stress tensor has been identified to lead to interesting and distinct signatures. For the purposes of our holographic interpretation however, we note that the deformation of the holographic CFT by the $T^2$ operator is the reason for this quadratic correction to the stress energy appearing. This reveals that the $T^2$ deformation provides the precise sense in which the holographic CFT, when brought to the Planck brane is cutoff.

\subsection{Comparison to alternative proposals}

Our interpretation differs from that of \cite{Shiromizu_2002_2} and \cite{Padilla_2006} where the bulk Weyl tensor term is kept and interpreted as the CFT stress energy tensor. We cannot quite incorporate this contribution into our interpretation because then we would need for there to be a term quadratic in $E^{\mu}_a$ appearing on the RHS of \eqref{eq:brane_efe}. 

Furthermore, in the context of the $T^2$ deformation, at any order in $\lambda$, we demand the covariant conservation of the stress tensor, i.e.
\begin{equation}
    \nabla_{\mu}T^{\mu}_a = 0
\end{equation}
which, when expanded to leading order in the deformation parameter, (i.e. in the regime of $\lambda\sim \epsilon$) gives 
\begin{equation*}
    \nabla_{\mu}T^{\mu}_a =0= \nabla_{\mu}\bigg( T^{\mu}_{CFT\,a} + \frac{\epsilon(D-1)}{2} \bigg(\bigg(T^{\mu}_{CFT\,d}f^{d}_{\sigma}T^{\sigma}_{CFT\,a}-\frac{1}{D-1}(T^{\rho}_{CFT\,d} f^{d}_{\rho})T^{\mu}_{CFT\,a}\bigg) -\end{equation*}\begin{equation} - \frac{f^\mu_a}{2}K^{cd}_{\alpha\beta}T^{\alpha}_{CFT\,c}T^{\beta}_{CFT\,d}\bigg)\bigg).
\end{equation}

However, the presence of the bulk Weyl tensor contribution leads to the equation
\begin{equation}
    \nabla_{\mu}T^{\mu}_a = \nabla_{\mu} E^{\mu}_a. 
\end{equation}
In other words, the presence of this contribution signals a violation of the covariant conservation of the stress tensor. In most studies of this topic, the divergence of $E^{\mu}_a$ is set to cancel the divergence of the quadratic contribution to what we call $T^{\mu}_a$, while the divergence of linear term in $T^{\mu}_a$ is assumed to vanish.

\section{Conclusions and Outlook}
\label{sec:conclusions}
In conclusion, we have shown how coupling a large N holographic CFT to the quadratic dRGT mass term is equivalent to deforming it by the $T^2$ operator. This deformation triggers a flow that mirrors the motion of the holographic boundary into the bulk in AdS/CFT. We have also showed that in the limit of small deformation parameter, the deformed CFT action provides exactly the right matter stress energy that sources the gravitational equations on the Planck brane in the context of the Randall-Sundrum II scenario. 

It is interesting to note that a slight generalization of the $T^2$ deformation, to include the boundary cosmological constant was identified as playing a key role in realizing the so called dS/dS scenario \cite{dSdSTTbar,Alishahiha:2004md}. In fact, the RS II scenario played a key role in the construction of this scenario, so it would be interesting to see if we can linearize the $T^2+\Lambda_d$ deformation and obtain the stress tensor appearing in the gravitational equations on a supercritical braneworld. 

It would also be interesting to see what modification of the dRGT mass term leads to the full non linear $T^2+\Lambda_d$ deformation. In two dimensions, it is known how this deformation provides a means to count the microstates that account for the Gibbons--Hawking entropy of the cosmic horizon \cite{dSMicro,Shyam_2022}. It would be very interesting to understand the higher dimensional generalization of this result from the lens of the formalism presented in this article. We should note however that unlike in $d=2$, we don't have the same integrability properties of the $T\bar{T}$ deformation and we have very little control beyond the large N limit, so the micro-state counting story is expected to be more complicated. 

Another important avenue for generalization is the case where there are other matter fields present in the theory. These fields are known to lead to additional double trace deformations identified in \cite{Hartman:2018tkw}. It would be interesting to have the effect of those deformations also arise from some quadratic terms in the effective action involving auxiliary fields. The deconstruction ansatz has been applied to (linearized) supergravity theories as well (see \cite{Ondo:2017kea} and references therein). It would be interesting to see if those mass terms can lead to a supersymmetric generalization of the $T^2$ deformation. 

Finally, it is well known that 11 dimensional supergravity compactified on $S^1/\mathbb{Z}$ produces Type I or Heterotic string theory in 10 dimensions \cite{Horava_1996_1,Horava_1996}. This bulk-boundary relationship is reminiscent of the Randall-Sundrum I scenario except with zero cosmological constant. It would be very interesting to see if there is some role that the $T^2$ (or related) deformations play in this context.

\section*{Acknowledgements} 
We thank Eva Silverstein and Gonzalo Torroba for comments on the draft of this article. 

V.S. is supported by the Branco Weiss Fellowship - Society in Science, administered by the ETH Zurich.

\bibliographystyle{utphys}
\bibliography{refs}

\end{document}